\documentclass[prd,reprint,twocolumn,showpacs,preprintnumbers,amsmath,amssymb,superscriptaddress,tightenlines,floatfix]{revtex4-1}
\usepackage{url}
\usepackage{bm}
\usepackage{graphicx}% Include figure files
\usepackage{epsfig}% Include figure files
\usepackage{rotating}% Include figure files
\usepackage{dcolumn}% Align table columns on decimal point
\usepackage{xcolor}
\usepackage{slashed}
\usepackage{threeparttable}
\usepackage[colorlinks=true
,urlcolor=blue
,anchorcolor=blue
,citecolor=blue
,filecolor=blue
,linkcolor=blue
,menucolor=blue
,pagecolor=blue
,linktocpage=true
,pdfproducer=medialab
,pdfa=true
]{hyperref}

\DeclareMathOperator{\re}{Re}

\newcommand{\beq}{\begin{eqnarray}} 
\newcommand{\eeq}{\end{eqnarray}} 
\begin{document}
%%%%%%%%%%%%%%%%%%%%%%%%%%%%%%%%%%%%%%%%%%%%%%%%%%%%%%%%%%%%%%%%%%%%%%%%%%%%%%%

\preprint{IPPP/16/14}
\title{Heavy neutrino impact on the triple Higgs coupling}

\author{J. Baglio}
\email{julien.baglio@uni-tuebingen.de}
\affiliation{Institut f\"{u}r Theoretische Physik, Eberhard Karls
  Universit\"{a}t T\"{u}bingen, Auf der Morgenstelle 14, D-72076
  T\"{u}bingen, Germany.}

\author{C. Weiland}
\email{cedric.weiland@durham.ac.uk}
\affiliation{Institute for Particle Physics Phenomenology, Department
  of Physics, Durham University, South Road, Durham DH1 3LE,
  United~Kingdom.}

\begin{abstract} 
We present the first calculation of the one-loop corrections to the
triple Higgs coupling in the framework of a simplified 3+1 Dirac
neutrino model, that is three light neutrinos plus one heavy neutrino
embedded in the Standard Model (SM). The triple Higgs coupling is a
key parameter of the scalar potential triggering the electroweak
symmetry-breaking mechanism in the SM. The impact of the heavy
neutrino can be as large as $+20\%$ to $+30\%$ for parameter points
allowed by the current experimental constraints depending on the
tightness of the perturbative bound. This can be probed at the
high-luminosity LHC, at future electron-positron colliders and at the
Future Circular Collider in hadron-hadron mode, an envisioned 100 TeV
$pp$ machine. Our calculation, being done in the mass basis, can be
extended to any model using the neutrino portal. In addition, the effects
that we have calculated are expected to be enhanced if additional
heavy fermions with large Yukawa couplings are included, as in
low-scale seesaw mechanisms.
\end{abstract} 

\maketitle

\section{INTRODUCTION}

The biggest highlight of the 7--8 TeV run of the
Large Hadron Collider (LHC) at CERN is the discovery in 2012 of a
Higgs boson with a mass of
around 125~GeV~\cite{Aad:2012tfa,Chatrchyan:2012ufa}. The
Higgs boson is the remnant of the electroweak symmetry-breaking (EWSB)
mechanism~\cite{Higgs:1964ia,Englert:1964et,Higgs:1964pj,Guralnik:1964eu}
that gives their masses to the other fundamental particles and
unitarizes the scattering of weak
bosons~\cite{Cornwall:1973tb,LlewellynSmith:1973yud}. While data
agrees well with Standard Model (SM) expectations, open questions
remain and the measure of the triple Higgs coupling would ultimately
test the electroweak sector of the SM by allowing the reconstruction
of the scalar potential. It is one of the main goals of the
high-luminosity run of the LHC (HL-LHC) and of future colliders, as it
is central to ongoing case studies; for reviews, see
e.g. Refs.~\cite{Baglio:2014aka,Baglio:2015wcg}.

Neutrino oscillations form the only confirmed phenomenon that
absolutely calls for a particle physics
explanation~\cite{Fukuda:1998mi} and an extension of the SM. One of
the simplest possibilities to explain the nonzero neutrino masses and
mixing is to add fermionic gauge singlets that will play the role of
right-handed neutrinos. The addition of these heavy sterile neutrinos
is quite generic, appearing in the type I
seesaw~\cite{Minkowski:1977sc,Ramond:1979py,GellMann:1980vs,Yanagida:1979as,Mohapatra:1979ia,Schechter:1980gr,Schechter:1981cv}
and its
variants~\cite{Mohapatra:1986aw,Mohapatra:1986bd,Bernabeu:1987gr,Pilaftsis:1991ug,Ilakovac:1994kj,Akhmedov:1995ip,Akhmedov:1995vm,Barr:2003nn,Malinsky:2005bi},
for example. In particular, these heavy sterile neutrinos behave like
pseudo-Dirac fermions in low-scale seesaw mechanisms that rely on an
approximately conserved lepton number. Here, we will consider a
simplified 3+1 model where the SM is phenomenologically modified to
account for 3 light massive neutrinos and one heavy sterile neutrino,
all of them being Dirac fermions. This will capture the beyond-the-SM
(BSM) effects that we want to analyze, effects that we expect to arise
in all models where the neutrino portal connects the SM with new
physics be it dark matter or hidden sectors.

In this article we study the impact of this extended neutrino sector on
the Higgs sector and in particular on the triple Higgs coupling,
proposing the latter as a new observable for neutrino physics and
showing that it can constrain heavy neutrinos in a mass regime
difficult to probe otherwise. We calculate for the first time the full
one-loop corrections to the triple Higgs coupling in a simplified 3+1
model with three light Dirac neutrinos, identified with those of the
SM, plus one heavy Dirac neutrino. We describe the analytical setup of
our calculation, the model as well as the theoretical and experimental
constraints considered before presenting the numerical results. For
parameter points that are allowed by the constraints we find effects
that can be probed at the HL-LHC, at future electron--positron
colliders and at the Future Circular Collider in hadron-hadron mode
(FCC-hh), a potential 100 TeV $pp$ collider following the LHC. These
large deviations are entirely due to quantum corrections, contrarily
to supersymmetric or composite models where tree-level corrections are
dominant, leading to a new type of scenario for the study of the
triple Higgs coupling. Moreover, larger deviations are expected in UV
complete models where more heavy neutrinos are included, as in the
case of low-scale seesaw mechanisms.

\section{CALCULATION SETUP}

We start with the SM scalar potential,
\begin{align}
V(\Phi) = & -\mu_{}^2 |\Phi|_{}^2 + \lambda |\Phi|_{}^4\,,
\end{align}
with
\begin{align}
\Phi = & \frac{1}{\sqrt{2}}
         \left(\begin{matrix} \sqrt{2} G_{}^+\\v+H+\imath
                 G_{}^0\end{matrix}\right)\,.
\end{align}
Here, $H$ is the Higgs boson, $G_{}^0$ the neutral Goldstone boson, $G_{}^\pm$
the charged Goldstone boson and $v\simeq 246$ GeV the vacuum
expectation value. We can define the Higgs tadpole $t_H^{}$,
the Higgs mass $M_H^{}$ and the triple Higgs coupling $\lambda_{HHH}^{}$
as the linear, quadratic and cubic term of the scalar potential,
respectively, in terms of the field $H$. At tree level, $t_H^{}=0$ and
the triple Higgs coupling is
\begin{align}
\lambda_{}^{0} = - \frac{3 M_H^2}{v}.
\end{align}

For the one-loop corrections to the triple Higgs coupling, our set of
input parameters that need to be renormalized in the on-shell (OS)
scheme is the following:
\begin{align}
M_H^{}, M_W^{}, M_Z^{}, e, t_H^{}.
\end{align}
We require that we have no tadpoles at one loop:
\begin{equation}
t_H^{(1)} + \delta t_H^{} = 0 \Rightarrow \delta t_H^{} = - t_H^{(1)},
\end{equation}
with $t_H^{(1)}$ being the one-loop diagrams for $t_H^{}$. We also
renormalize the Higgs wave function in the OS scheme. The full
renormalized one-loop triple Higgs coupling is then $\lambda_{HHH}^{1r}
= \lambda_{}^0 + \lambda_{HHH}^{(1)} + \delta\lambda_{HHH}^{}$ with
\begin{align}
  \frac{\delta\lambda_{HHH}^{}}{\lambda_{}^0}  =\, & \frac32 \delta Z_H^{} +
 \delta t_H^{} \frac{e}{2 M_W^{}\sin\theta_W^{} M_H^2} +\delta Z_e^{} +
\frac{\delta M_H^2}{M_H^2} \nonumber\\ 
&  - \frac{\delta M_W^2}{2 M_W^2} +
\frac12 \frac{\cos_{}^2\theta_W^{}}{\sin_{}^2\theta_W^{}} \left( \frac{\delta
 M_W^2}{M_W^2}-\frac{\delta M_Z^2}{M_Z^2}\right)\,,
\end{align}
where $\delta Z_H^{}, \delta Z_e^{}, \delta M_H^2, \delta M_W^2, \delta
M_Z^2$ are the counterterms of the Higgs fields, the electrical
charge, the Higgs mass and the $W_{}^\pm/Z$ weak boson masses,
respectively, $\theta_W^{}$ is the weak mixing angle and
$\lambda_{HHH}^{(1)}$ stands for the one-loop diagrams of the process
$H_{}^{*}\to H H$. We define, for the analysis of the results,
\begin{align}
  \Delta_{}^{(1)} \lambda_{HHH}^{} & =
                               \frac{1}{\lambda_{}^{0}}\left(\lambda_{HHH}^{1r}
                               -\lambda_{}^0\right)\,,\,\nonumber\\
  \Delta_{}^{\rm BSM} & = \frac{1}{\lambda_{HHH}^{1r,{\rm
                     SM}}}\left(\lambda_{HHH}^{1r,{\rm full}}
                     -\lambda_{HHH}^{1r,{\rm SM}}\right)\,.
                     \label{eq:definedelta}
\end{align}

In order to be independent of the light fermion masses, we use the
following condition for the electric charge
renormalization~\cite{Denner:1991kt,Nhung:2013lpa},
\begin{align}
\delta Z_e^{}  =
  \frac{\sin\theta_W^{}}{\cos\theta_W^{}}\frac{\re\Sigma^T_{\gamma
  Z}(0)}{M_Z^2} - \frac{\re\Sigma^T_{\gamma
  \gamma}(M_Z^2)}{M_Z^2}\,,
\end{align}
where $\Sigma_{XY}^{}$ stands for the self-energy of the process $X\to
Y$.

\section{\boldmath SIMPLIFIED $3+1$ MODEL AND CONSTRAINTS}

In order to 
illustrate the effect of a heavy neutrino on the triple Higgs coupling
we introduce a simplified model that includes 3 light neutrinos and an
extra heavy neutrino. All of them are Dirac fermions and the heavy
neutrino couples to the SM particles through its mixing with SM
fields. This reproduces the behavior of heavy sterile neutrinos present in
seesaw extensions of the SM with approximately conserved lepton number for
example~\cite{Mohapatra:1986aw,Mohapatra:1986bd,Bernabeu:1987gr,Akhmedov:1995ip,Akhmedov:1995vm,Barr:2003nn,Malinsky:2005bi}. In the mass
  basis, the relevant couplings between neutrinos and SM bosons are
  given by
\begin{align}
 \mathcal{L}\ni & - \frac{g_2^{}}{\sqrt{2}}\bar \ell_i^{} \slashed{W}_{}^{-}
                  B_{i j}^{} P_L^{} n_j^{} + \mathrm{H.c.}\nonumber \\ 
   & -\frac{g_2^{}}{2\cos \theta_W^{}} \bar n_i^{} \slashed Z C_{i
     j}^{} P_L^{} n_j^{} \nonumber \\
   & - \frac{g_2^{}}{\sqrt{2} M_W^{}} \bar \ell_i^{} G_{}^- B_{i j}^{}
     (m_{\ell_i}^{} P_L^{}  - m_{n_j}^{} P_R^{} ) n_j^{} +
     \mathrm{H.c.}\nonumber \\ 
   & - \frac{g_2^{}}{2 M_W^{}} \bar n_i^{} C_{ij}^{} H (m_{n_i}^{}
     P_L^{} + m_{n_j}^{} P_R^{}) n_j^{} \nonumber \\ 
   & + \frac{ \imath g_2^{}}{2 M_W^{}} \bar n_i^{} C_{ij}^{} G_{}^0 (-
     m_{n_i}^{} P_L^{} + m_{n_j}^{} P_R^{}) n_j^{}\,,
\end{align}
where $\ell_i^{}$ are the charged leptons of mass $m_{e/\mu/\tau}^{}$,
$n_i^{}$ the Dirac neutrinos of mass $m_{1\cdots 4}^{}$, $g_2^{}$ is the
$\mathrm{SU}(2)$ coupling constant, and $B$ and $C$ are $4\times4$
mixing matrices. In particular, $B$ and $C$ are defined as
\begin{align}
 B=&R_{34}^{} R_{24}^{} R_{14}^{} \tilde U_{PMNS}^{}\,, \\ 
 C_{ij}^{}=&\sum_{k=1}^{3} B^*_{ki} B_{kj}^{}\,,
 \end{align}
with rotation matrices $R_{34}^{}$, $R_{24}^{}$ and $R_{14}^{}$ such
as
\begin{align}
 R_{14}^{}=\left(\begin{array}{cccc}
  \cos{\theta_{14}^{}} & 0 & 0 & \sin{\theta_{14}^{}} \\
  0 & 1 & 0 & 0 \\
  0 & 0 & 1 & 0 \\
  -\sin{\theta_{14}^{}} & 0 & 0 & \cos{\theta_{14}^{}}
\end{array}\right)\,,
\end{align}
and $\tilde U_{PMNS}^{}$ the block-diagonal matrix
\begin{equation}
 \tilde U_{PMNS}^{} = \left(\begin{array}{cc}
  U_{PMNS}^{} & 0
  \\ 0 & 1 
\end{array}\right)\,,
\end{equation}
where $U_{PMNS}^{}$~\cite{Pontecorvo:1957cp,Maki:1962mu}
corresponds to the best-fit point for a normal hierarchy
in~\cite{Gonzalez-Garcia:2014bfa} with $\delta_{CP}^{}=0$. We have also
chosen the three light neutrinos to be degenerate with
$m_{n_1/n_2/n_3}^{}=1\,\mathrm{eV}$ in agreement with the results of the
Mainz and Troitsk experiments~\cite{Kraus:2004zw,Aseev:2011dq}. Since
their small masses translate into small couplings to the Higgs boson,
we expect the corrections to the triple Higgs coupling from the light
neutrinos to be irrelevant. As a consequence, we have not varied the
light neutrino parameters in this work.

Other experimental and theoretical constraints that apply to the heavy
sterile neutrino have to be taken into account as well. Direct
searches at the LHC are not as constraining~\cite{Deppisch:2015qwa,Das:2015toa} as
indirect constraints yet and should be even less constraining than
claimed in Ref.~\cite{Deppisch:2015qwa,Das:2015toa} since the production cross sections
were overestimated according to Ref.~\cite{Degrande:2016aje}. Global
fits to various observables have been performed
recently~\cite{Antusch:2014woa,Deppisch:2015qwa,Antusch:2015mia,deGouvea:2015euy},
pointing to electroweak precision observables as the most constraining
ones above the Higgs mass. We use here the constraints from the global
fit performed in~\cite{delAguila:2008pw,deBlas:2013gla},
\begin{align}
 B_{e4}^{}\,\leq\, & \, 0.041\,, \nonumber \\
 B_{\mu4}^{}\,\leq\, & \, 0.030\,, \nonumber \\
 B_{\tau4}^{}\,\leq\, & \, 0.087 \label{maxUtau}\,,
\end{align}
at the $95\%$~C.L. From the theoretical point of view, we require the
loop expansion to remain perturbative, applying either a loose (tight)
bound of
\begin{equation}
 \left(\frac{\mathrm{max}|C_{i4}^{}|\,g_2^{}\,  m_{n_4}^{}}{2 M_W^{}}
 \right)_{}^3 < 16 \pi\,(2\pi)\,.
\end{equation}
The tight bound is roughly equivalent to the bound that comes from a
two-loop analysis of the perturbativity of the
SM presented in~\cite{Durand:1993vn,Nierste:1995zx}. Using the largest
$C_{i4}^{}$ in agreement with Eq.~(\ref{maxUtau}), these translate
into upper limits on the heavy neutrino mass of $m_{n_4}^{}=14.3$~TeV
and $m_{n_4}^{}=7.2$~TeV, respectively. However, the decay width of
the heavy neutrino grows as $m_{n_4}^3$. In order for the quantum
state to be a definite particle, we also require
$\Gamma_{n_4}^{}\leq0.6\,m_{n_4}^{}$, which limits the upper value of
$m_{n_4}^{}$ to approximately 9~TeV for $B_{\tau4}^{}=0.087$. These
limits depend on the values chosen for the $B_{i4}^{}$ and a decreased
mixing increases the upper limit on $m_{n_4}^{}$.

\begin{figure}[!b]
 \centering
 \includegraphics[scale=0.75]{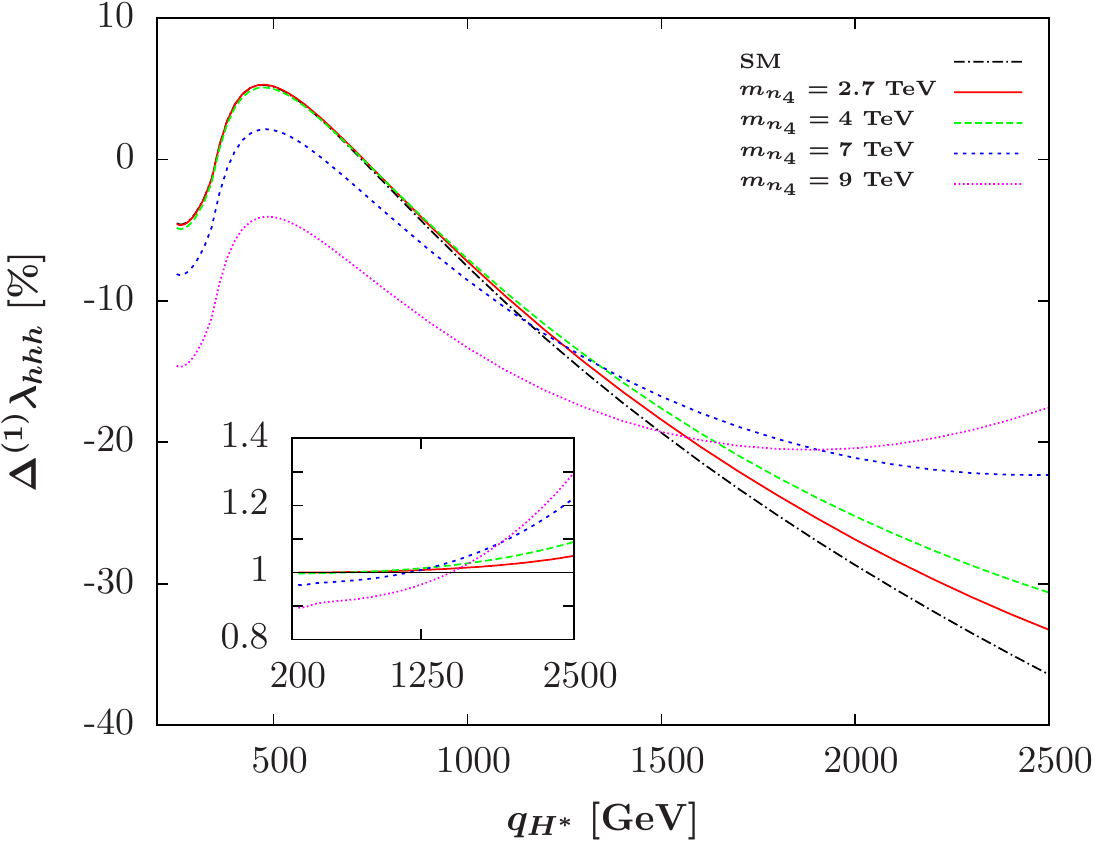}
\caption[]{One-loop corrections to the triple Higgs coupling
  $\lambda_{HHH}^{}$ (in \%) as a function of the momentum $q_{H^*}^{}$
  of the splitting $H^*(q_{H^*}^{})\to HH$ (in GeV), for several values of
  the neutrino mass parameter $m_{n_4}$ and at a constant $B_{\tau 4}^{} =
  0.087$. The ratio of the genuine BSM contribution to the triple
  Higgs coupling with respect to the one-loop SM contribution is shown
  in the insert.}
\label{hhh_dirac}
\end{figure}

\section{RESULTS}

For the numerical evaluation of the one-loop corrections,
the SM parameters are chosen as

\begin{align}
  m_t^{} = 173.5~\text{GeV}\,,\,\, M_H^{} = 125~\text{GeV}\,,\,\,
  M_W^{} = 80.385~\text{GeV}\,,\nonumber
\end{align}
%\vspace{-9mm}
\vspace{-10mm}
\begin{align}
  M_Z^{} = 91.1876~\text{GeV}\,,\,\, \alpha_{}^{-1} = 127.934\,.
\end{align}
The new neutrino contributions to the one-loop diagrams of
$\lambda^{(1)}_{HHH}$ are
\begin{widetext}
\begin{align}
  \lambda^{(1,\nu)}_{HHH} = & -\frac{\alpha\sqrt{4\pi\alpha}}{32 \pi M_W^3 s_W^3} 
  \sum _{j,k,l=1}^4 \left(C_{j k}^* C_{k l}^* C_{l j}^*+C_{j l}^* C_{l
 k}^* C_{k j}^* \right) \left[m_{n_j}^2
   m_{n_k}^2\Big(4 q_{H^*}^2 C_1^{}+(4 M_H^2+q_{H^*}^2) C_2^{}\Big) \right. \nonumber\\
 & +m_{n_l}^2\left(q_{H^*}^2 C_1^{} (5 m_{n_j}^2+3 m_{n_k}^2)+
  C_2^{} \Big(2 m_{n_j}^2 (M_H^2+q_{H^*}^2) +m_{n_k}^2 (2 M_H^2+q_{H^*}^2)
   \Big)\right) \nonumber\\
 & \left. +4 B_0^{} \left(m_{n_k}^2 m_{n_l}^2+ m_{n_l}^2 m_{n_j}^2 +
   m_{n_j}^2 m_{n_k}^2\right) +C_0^{}
   m_{n_j}^2 \left( m_{n_l}^2 (q_{H^*}^2+4 m_{n_j}^2) 
 +m_{n_k}^2 (q_{H^*}^2+4 m_{n_j}^2+8 m_{n_l}^2)\right)\right]\,,
\label{eq:1loopneutrinos}
\end{align}
\end{widetext}
with $B_0^{}$, $C_0^{}$, $C_1^{}$ and $C_2^{}$ the scalar and
tensor integrals~\cite{'tHooft:1978xw,Passarino:1978jh}
\begin{align}
B_0^{} & \equiv  B_0^{}\left(M_H^2,m_{n_k}^2,m_{n_l}^2\right)\,,\nonumber\\
C_0^{} & \equiv  C_0^{}\left(q_{H^*}^2,M_H^2,M_H^2,
 m_{n_j}^2,m_{n_k}^2,m_{n_l}^2\right)\,,\nonumber\\
C_{1/2}^{} &  \equiv  C_{1/2}^{}\left(q_{H^*}^2,M_H^2,M_H^2,
m_{n_j}^2,m_{n_k}^2,m_{n_l}^2\right)\,.
\end{align}
This comes in addition to the SM corrections that we have recalculated
and found to agree with the literature, see, for example,
Ref.~\cite{Arhrib:2015hoa} and references therein.

\begin{figure*}[!t]
   \centering
   \includegraphics[scale=0.71]{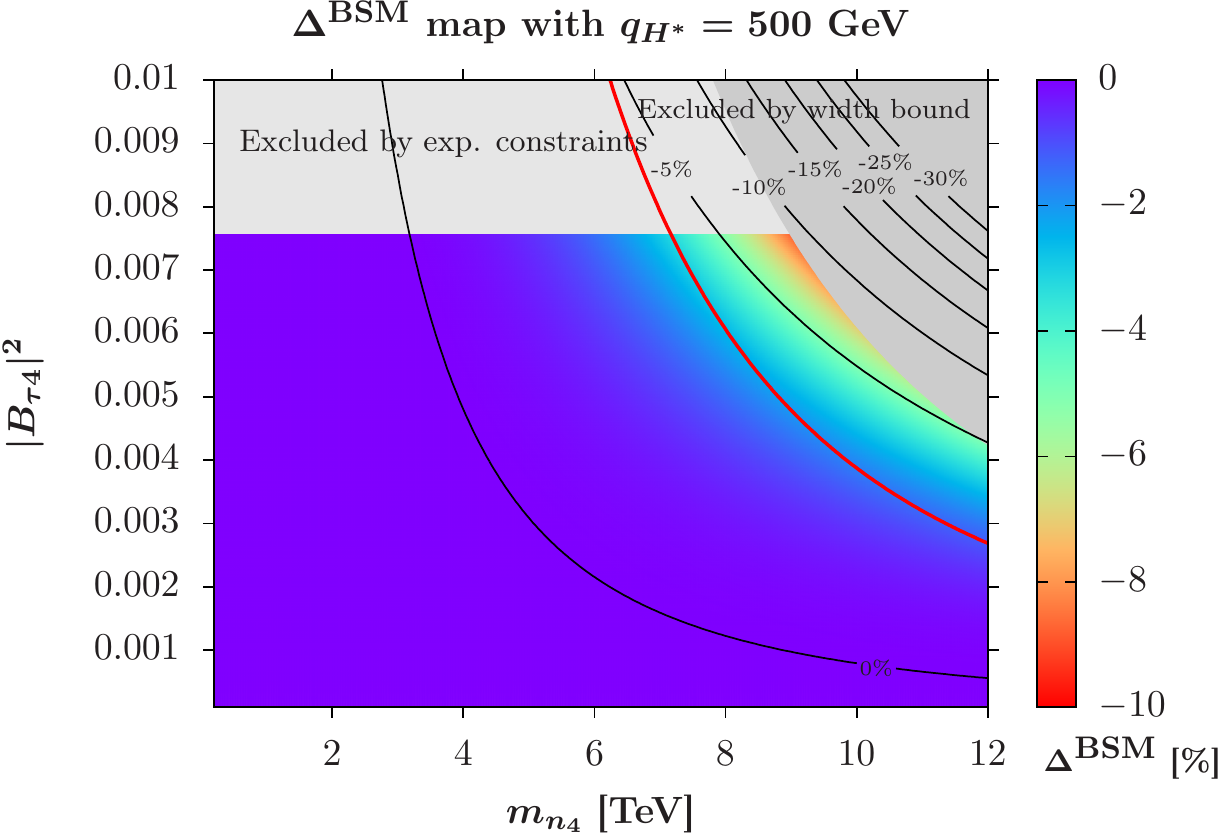}
   \includegraphics[scale=0.71]{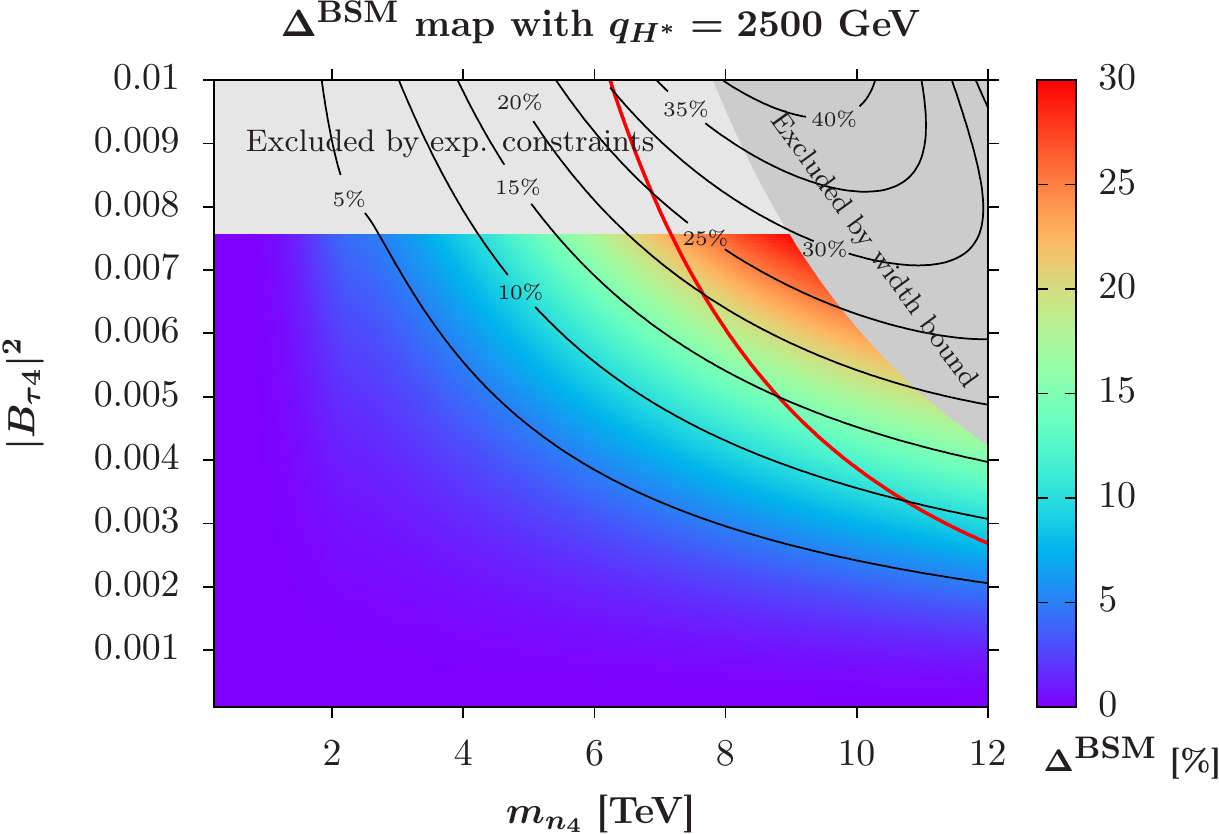}
   \caption{Contour maps of the neutrino corrections $\Delta_{}^{\rm
       BSM}$ to the triple Higgs coupling $\lambda_{HHH}^{}$ (in percent)
     as a function of the two neutrino parameters $|B_{\tau 4}^{}|_{}^2$ and
     $m_{n_4}^{}$ (in TeV), at a fixed off-shell Higgs momentum $q_{H^*}^{} =
     500$ GeV (left) and $q_{H^*}^{}=2500$ GeV (right). The other heavy
     neutrino mixing parameters are set to zero. The light gray area
     is excluded by the experimental constraints and the darker gray
     area is excluded from having $\Gamma_{n_4}^{}>0.6\,m_{n_4}^{}$ while
     the red line corresponds to the tight perturbativity bound.
     \label{fig:hhh-scan}}
 \end{figure*}

In Fig.~\ref{hhh_dirac}, we present the full one-loop correction,
including the neutrinos contribution, as a function of $q_{H^*}^{}$
where the genuine BSM effects are depicted in the insert as ratio over
the SM one-loop result. Here $q_{H^*}^{}$ is the momentum of the
initial off-shell Higgs boson in the splitting $H(q_{H^*}^{})\to
HH$. We have varied $m_{n_4}^{}$ while keeping $B_{\tau4}^{} = 0.087$
fixed at its maximum allowed value given in Eq.~(\ref{maxUtau}). The
other neutrino mixing terms are zero.

The choice $m_{n_4}^{} = 2.7$ TeV corresponds to
the case where the heavy neutrino effective coupling to the Higgs
boson is equal to the top-quark Yukawa coupling,
i.e.~$\mathrm{max}(C_{i4}) m_4 \sim m_t$, while $m_{n_4}^{} = 7$ TeV
and $m_{n_4}^{} = 9$ TeV respectively correspond to the tight
perturbativity and the width bounds. As can be seen, all scenarios
lead to sizable one-loop corrections at low momentum. The largest
positive one-loop corrections in the SM are at $q_{H^*}^{}\simeq 500$
GeV, where the BSM contribution tends to decrease them to $-3\%$ at
$m_{n_4}^{} = 7$ TeV and $-9\%$ at $m_{n_4}^{}=9$ TeV. The most
interesting effects come at larger momentum where the deviation from
the SM increases with larger $m_{n_4}^{}$, reaching $+22\%$ at
$q_{H^*}^{} = 2500$ GeV for $m_{n_4}^{} = 7$ TeV and even $+30\%$ for
$m_{n_4}^{}=9$ TeV. 

To get an idea of the dependence of the heavy neutrino effect on the
model parameters, we have scanned in Fig.~\ref{fig:hhh-scan} the
parameter space $(m_{n_4}^{},|B_{\tau 4}^{}|_{}^2)$ while fixing the
remaining parameters $B_{e 4}^{}=B_{\mu 4}^{} = 0$. $\Delta_{}^{\rm
  BSM}$ is the percentage correction of the neutrino effects over the
one-loop SM triple Higgs coupling defined in
Eq.~(\ref{eq:definedelta}). The off-shell Higgs momentum has been
fixed to the most interesting values identified in
Fig.~\ref{hhh_dirac}, $q_{H^*}^{} = 500/2500$ GeV and the experimental
and theoretical bounds are included as gray areas. When the heavy
neutrino mass and mixing are small, its contribution vanishes as
expected. As is seen in Eq.~(\ref{eq:1loopneutrinos}), the BSM correction
exhibits a strong dependence on the heavy neutrino mass, as terms of
order $\mathcal{O}(m_{n_4}^4)$ and $\mathcal{O}(m_{n_4}^6)$ arise and
they increase $|\Delta_{}^{\rm BSM}|$ gradually to reach the maximum
allowed deviations $\Delta_{}^{\rm BSM} \simeq -10\%$ at
$q_{H^*}^{}=500$ GeV and $\Delta_{}^{\rm BSM} \simeq +30\%$ at
$q_{H^*}^{}=2500$ GeV. Similar contour maps are found when one trades
$B_{\tau 4}^{}$ against either $B_{e 4}^{}$ or $B_{\mu 4}^{}$.

Given the projected sensitivity of around $50\%$ per experiment
on the measure of $\lambda_{HHH}^{}$ at the
HL-LHC~\cite{CMS-PAS-FTR-15-002}, which translates into a sensitivity
of $35\%$ when statistically combining ATLAS and CMS, the effect of
the heavy neutrino may already be probed at the HL-LHC in the case of
the maximal value of $m_{n_4}^{} = 9$ TeV. The International Linear
Collider (ILC), one of the future potential electron-positron
colliders, at $500$~GeV would reach a precision of $27\%$ with
4~ab$_{}^{-1}$ while at $1$~TeV with 5~ab$_{}^{-1}$ the projected
sensitivity improves to $10\%$~\cite{Fujii:2015jha}, making the
effects clearly visible. At the FCC-hh the effects become important
enough to constrain the heavy neutrino mass and mixing, as the
projected statistical precision on $\lambda_{HHH}^{}$ at the FCC-hh
with 3 ab$_{}^{-1}$ is expected to be $13\%$ per experiment using only
the $b\bar{b}\gamma\gamma$~\cite{He:2015spf}. Combining the two
experiments and using the other search channels, it is reasonable to
expect a $\sim 5\%$ sensitivity which is the target sensitivity of the
FCC for this observable. The $+ 22\%$ increase predicted at
high $q_{H^*}^{}$ with a conservative perturbativity limit is 4 times
the projected sensitivity. Clearly, measuring the neutrino effect or
constraining neutrino models in a region hard to probe otherwise
becomes possible.

\section{SUMMARY AND OUTLOOK}

In this article we have investigated the
one-loop effects of a heavy neutrino on the triple Higgs coupling in a
simplified model that accounts for the light neutrino masses and
mixing and contains one extra heavy neutrino. After taking into
account the experimental constraints we have found that the maximum
effect can be as large as a 30\% increase over the SM one-loop
effects, independently of specific flavor structures. The effect is
of the order of the projected experimental accuracy on the measure of
$\lambda_{HHH}^{}$ at the HL-LHC, and if a tighter perturbative bound
is used on the 3+1 model, it is still possible to have a +20\%
deviation that is 2 times the projected experimental accuracy at the
ILC at $1$~TeV with 5~ab$_{}^{-1}$ and 4 times the one at the
FCC-hh with 3~ab$_{}^{-1}$, thus clearly visible. This is the first
time the effects of an extended neutrino sector on the triple Higgs
coupling have been investigated, demonstrating that it provides a
novel way of probing neutrino mass models in a regime otherwise
difficult to access. This provides an extra motivation to the
experimental measurement of this coupling. It should be noted that the
calculation can be extended to all models using the neutrino portal,
like dark matter models, and to cases with more heavy neutrinos where
the effects can be enhanced. This is confirmed by our preliminary
study in the inverse seesaw~\cite{inpreparation}.\bigskip

\section*{ACKNOWLEDGMENTS}

The authors thank Silvia Pascoli and Carlos Tamarit for stimulating
discussions on $\mathrm{SU}(2)$ breaking and renormalization and
Oliver Fischer for bringing our attention to the ILC
sensitivity. Discussions with Richard Ruiz are also
acknowledged. Peter Ballett, Silvia Pascoli and Richard Ruiz are also
thanked for their comments on this manuscript. A special thanks is
dedicated to Margarete M\"{u}hlleitner for having given a seminar that
initiated the project. J.~B. thanks the Institute for Particle Physics
Phenomenology for its hospitality in the last stages of the project
and acknowledges the support from the Institutional Strategy
of the University of T\"ubingen (DFG, ZUK 63) and from the DFG Grant
JA 1954/1. C.~W. receives financial support from the European Research
Council under the European Union's Seventh Framework Programme
(FP/2007-2013) / ERC Grant NuMass Agreement No.~617143 and partial
support from the EU Grant No. FP7 ITN INVISIBLES (Marie Curie Actions,
Grant No. PITN-GA-2011-289442) and from the European Union's Horizon
2020 research and innovation programme under the Marie
Sk\l{}odowska-Curie grant agreement No.~690575 and No.~674896.

\bibliography{hhh_neutrinos}

\end{document}